\documentclass[aps,pre,twocolumn,superscriptaddress]{revtex4}

\usepackage{graphicx,amssymb,amsmath}
\usepackage{color} 
\usepackage{xcolor}
\usepackage{multirow}

\newcommand{\be}{\begin{equation}}
\newcommand{\ee}{\end{equation}}
\newcommand{\bea}{\begin{eqnarray}}
\newcommand{\eea}{\end{eqnarray}}
\newcommand{\br}{\mathbf{r}}
\newcommand{\bt}{\mathbf{t}}

\def\eq#1{Eq.~(\ref{#1})}
\def\eqs#1#2{Eqs.~(\ref{#1},\ref{#2})}

\begin{document}

\title{Dependence of DNA persistence length on ionic strength of solutions with monovalent and divalent salts: a joint theory-experiment study}

\author{Anna\"el Brunet}
\affiliation{Institut de Pharmacologie et Biologie Structurale, UMR CNRS-UPS 5089, 31062 Toulouse, France} 
\affiliation{Universit\'e de Toulouse, UPS, Institut de Pharmacologie et Biologie Structurale, 31062 Toulouse, France}
\affiliation{Universit\'e de Toulouse, UPS, Laboratoire de Physique Th\'eorique (IRSAMC), 31062 Toulouse, France}
\affiliation{CNRS, Laboratoire de Physique Th\'eorique (IRSAMC), 31062 Toulouse, France}
\author{Catherine Tardin}
\author{Laurence Salom\'e}
\affiliation{Institut de Pharmacologie et Biologie Structurale, UMR CNRS-UPS 5089, 31062 Toulouse, France} 
\affiliation{Universit\'e de Toulouse, UPS, Institut de Pharmacologie et Biologie Structurale, 31062 Toulouse, France}
\author{Philippe Rousseau}
\affiliation{Universit\'e de Toulouse, UPS, LMGM (Laboratoire de
Microbiologie et G\'en\'etique Mol\'eculaires), 31062 Toulouse, France}
\affiliation{CNRS, LMGM, UMR CNRS-UPS 5100, 31062 Toulouse, France} 
\author{Nicolas Destainville}
\author{Manoel Manghi}
\affiliation{Universit\'e de Toulouse, UPS, Laboratoire de Physique Th\'eorique (IRSAMC), 31062 Toulouse, France}
\affiliation{CNRS, Laboratoire de Physique Th\'eorique (IRSAMC), 31062 Toulouse, France}

\date{\today}

\begin{abstract}
Using high-throughput Tethered Particle Motion single molecule experiments, the double-stranded DNA persistence length, $L_p$, is measured in solutions with Na$^+$ and Mg$^{2+}$ ions of various ionic strengths, $I$. Several theoretical equations for $L_p(I)$ are fitted to the experimental data, but no decisive theory is found which fits all the $L_p$ values for the two ion valencies. Properly extracted from the particle trajectory using simulations, $L_p$ varies from 30~nm to 55~nm, and is compared to previous experimental results. For the Na$^+$ only case, $L_p$ is an increasing concave function of $I^{-1}$, well fitted by Manning's electrostatic stretching approach, but not by classical Odjik-Skolnick-Fixman theories with or without counter-ion condensation. With added Mg$^{2+}$ ions, $L_p$ shows a marked decrease at low $I$, interpreted as an ion-ion correlation effect, with an almost linear law in $I^{-1}$, fitted by a proposed variational approach.
\end{abstract}
\maketitle

\section{Introduction}

Ions play a major role in the cell, for example by modifying the protein activity, inducing a voltage between the intra-cellular and extra-cellular matrix, or controlling the DNA packaging in viral capsids or in the nucleus. Complex mechanisms involving DNA such as its wrapping around histones or its denaturation will only be thoroughly understood when the role of mobile ions on the DNA conformation gets elucidated, DNA  being one of the most charged biopolymers found in nature.

The first quantitative experimental studies of DNA conformational properties as a function of salt concentration have been done in 1978 by Harrington~\cite{Harrington1978}, using flow birefringence (FB) experiments, to measure the DNA radius of gyration in dilute DNA solutions.
The DNA radius of gyration is intimately related to the DNA persistence length, namely the correlation length of the tangent-tangent correlation function 
\be
\langle \bt(s)\cdot\bt(0)\rangle=\exp\left(-s/L_p\right)
\label{WLCcorr}
\ee
where $\bt(s)$ is the unit vector tangent to the chain at the point of curvilinear index $s$. The persistence length, $L_p$, thus characterizes the chain stiffness at small length scales.  

Experimentally, the persistence length cannot be directly measured and the required procedure to extract it has remained a major issue since these first quantitative measurements~\cite{Harrington1978}.  Other optical techniques have then been used such as the transient electric birefringence (TEB)~\cite{Hagerman1981} or magnetic birefringence (MB)~\cite{Maret1983}, linear dichroism (LD)~\cite{Rizzo1981}, dynamic light scattering (DLS)~\cite{Borochov1981}, and force-stretching by optical tweezers (FOT)~\cite{Baumann1997} were used to estimate the variation of $L_p$ as a function of the ionic strength $I$.
In a recent paper, Savelyev~\cite{Savelyev2012} reviewed the available experimental data and showed that they could be divided into two groups, based on the distinct behaviours of $L_p$ found at high ionic strength. Indeed whereas the first group of experimental data~\cite{Harrington1978,Rizzo1981,Maret1983,Baumann1997} indicated a slow decrease of $L_p$ with increasing $I$, the second one~\cite{Kam1981,Manning1981,Borochov1981,Post1983,Sobel1991,Cairney1982} found a significant one.

Hence no global picture emerges yet from the literature. Many reasons can be put forward, such as the difficulty to estimate accurately the ionic strength in buffers, which is not simply equal to the added salt concentration, or the method of extraction of the persistence length from the experimental observables. Indeed, the extraction of the variations of $L_p$ with $I$ by using FB, MB and LD techniques is very sensitive to the optical arrangement between the electric field and the molecular axis, which is related to the tedious evaluation of the magnetic/optical anisotropy of a single base-pair. Moreover, in LD experiments, the mechanism of interaction between nucleic acids and the electric field depends on the polarizability of the ionic cloud surrounding the DNA, which therefore requires an additional modeling. In FOT experiments, the high DNA stretching induced by the force modifies the DNA structure and the organization of its ionic cloud. Finally in DLS experiments, the DNA hydrodynamic radius is estimated from the measure of the diffusion coefficient. Inferring the persistence length is not easy especially due to excluded volume effects.

Among the different theories developed to explain the variations of $L_p$ with the ionic strength, $I$, the most famous is the Odjik-Skolnick-Fixman (OSF)~\cite{Odijk1977,Skolnick1977} theory where $L_p$ is the sum of a bare, non-electrostatic persistence length and an electrostatic contribution scaling as $I^{-1}$. This theory accounts, at least qualitatively, for the fact that a rise in $I$ provokes an increasing screening of the repulsive phosphate ions of the backbone which leads to a more flexible DNA chain.
Taking into account the so-called Manning counter-ion condensation around the DNA at low $I$, the prefactor in front of $I^{-1}$, proportional to the square of the effective DNA charge, is lowered. All these experimental works have therefore been compared to these types of approaches, with no decisive conclusion~\cite{Savelyev2012}. More recently, Manning proposed a new theory~\cite{Manning2006}, and explicitly considers the electrostatic stretching force of the polyelectrolyte, which qualitatively fits some experiments~\cite{Baumann1997} or numerical results at high $I$~\cite{Savelyev2010,Savelyev2012}.

In this paper we reconsider the old issue, still under debate~\cite{Peters2010}, of the dependence of DNA conformation, at room temperature, with the ionic strength of the surrounding solution, using the recent single-molecule technique of high-throughput Tethered Particle Motion (HT-TPM)~\cite{Plenat2012,Brunet2015}.
We measure the persistence length of two DNA of 1201 and 2060~base-pairs (bp) for a large range of well controlled ionic strengths from $I=10^{-2}$ to 3~mol/L with Na$^+$ counter-ions and with or without Mg$^{2+}$ added counter-ions. In the first Section, we present the simple and well controlled HT-TPM experiments and their analysis. The next Section is devoted to the experimental results and the extraction of the persistence length from the HT-TPM amplitude of motion using numerical simulations. Our results are then compared to previous ones, and to the various existing theories as well as our detailed variational approach and an interpolation formula that fits all our experimental $L_p$. Finally our concluding remarks are given in the last Section.

\section{Experimental section}
\label{expsect}

\subsection{High-Throughput Tethered Particle Motion (HT-TPM) experimental procedure}
\begin{figure}[!t]
\includegraphics[width=0.8\columnwidth]{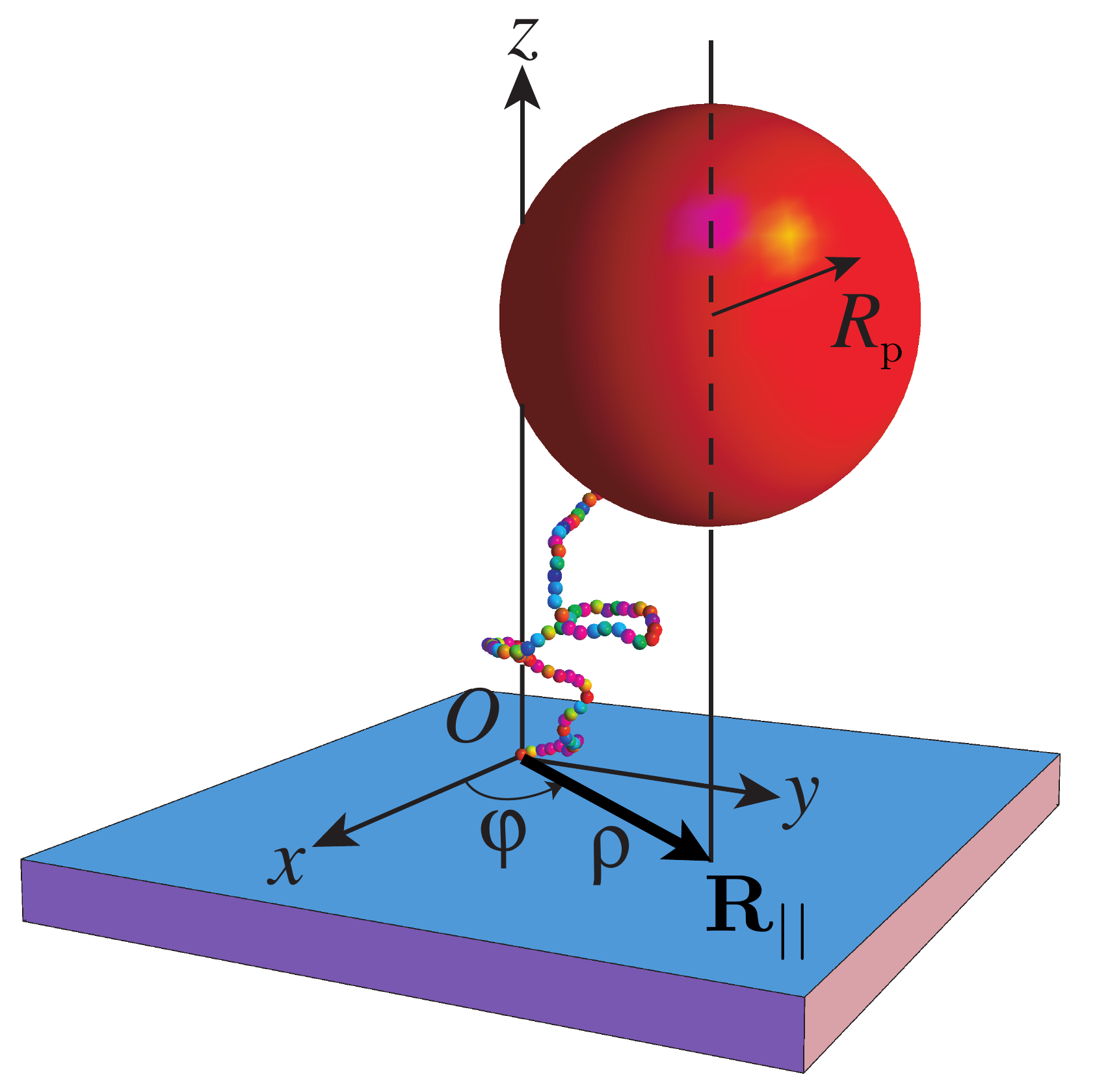}
\caption{\label{TPM} Principle of the Tethered Particle Motion experiment. The DNA is tethered at one end on the coverslip and labelled by a particle (radius $R_{\rm p}$) at the other end. The projected particle position $\mathbf{R}_{||}$ is tracked as a function of time. The DNA is model by a chain of beads in the simulations.}
\end{figure}

DNA molecules were produced by PCR amplification (oligonucleotides from Sigma-Aldrich) : Biot-F1201 5'-CTGGTGAGTACTCAACCAAG-3' and Dig-R1201 5'-CTACAATCCATGCCAACC-3' on pTOC1 plasmid and Biot-F2060 5'-CTGCAATGATACCGCGAGAC-3' and Dig-R2060 5'-TGACTTCCGCGTTTCCAGAC-3' on pBR322.

HT-TPM permits the simultaneous tracking of hundreds of single DNA molecules free to fluctuate in solution that are tethered to a coverslip at one end and labelled by a 300~nm particle at the other end (see Fig.~\ref{TPM}).
HT-TPM on chip assembly is performed as previously described in detail in Ref.~\cite{Brunet2015}. The anchoring of the DNA-particle complexes to the neutravidin (Invitrogen) printed sites was performed in  PBS buffer (Euromedex) supplemented with 1mg/mL of pluronic 127 (Sigma-Aldrich) and 0.1~mg/mL BSA (Sigma-Aldrich), noted T-BSA-Plu. 

A large range of buffers were then used to test the effect of both ion valency, by using monovalent Na$^+$ or divalent Mg$^{2+}$, and salt concentration, on the DNA conformations (Tables~1 and~2 in the Supporting Information). The first buffer, corresponding to zero salt added, and named Zero-salt-buffer in the following, is a phosphate buffer (KH$_2$PO$_4$ 1~mM, Na$_2$HPO$_4$ 3~mM, $p$H 7.4, pluronic F127 1~mg/mL). From this one, we added successively various concentrations of NaCl or MgCl$_2$ to obtain a large range of salt conditions (X-salt-buffer). Before starting the experiment, the flow cell was extensively rinsed with the Zero-salt-buffer ($\sim 100$ chamber volumes), let incubated during 1~h at room temperature, then rinsed again with $\sim 100$ chamber volumes of Zero-salt-buffer. The experiment started with a Zero-salt-buffer measurement, next the concentration in monovalent ions was progressively increased by addition of $\sim 100$ chamber volumes of X-salt-buffer.  Then the flow cell was extensively rinsed with the Zero-salt-buffer ($\sim 100$ chamber volumes), incubated during 4~min and rinsed again with by $\sim 100$  chamber volumes with the Zero-salt-buffer. A new Zero-salt-buffer measurements was performed. At last, the divalent ion concentration was progressively increased and new acquisitions were carried out. For all conditions, the acquisitions were performed at a controlled temperature equal to 25$^\circ$C, and a movie of 1~min were recorded and analyzed. We ensure the reliability of the experimental procedure by checking the agreement between the two values of the Zero-salt measurement obtained before the addition of monovalent ions and before the addition of divalent ions. Experiments were repeated on different days to ensure the reproducibility of our results.

The tethered particles of diameter 300~nm were visualized using a dark-field microscope (Axiovert 200, Zeiss) equipped with a $\times32$ objective and an additional x1.6 magnification lens and a temperature control system (Physitemp TS-4MPER). Images were acquired during 1~min at a frame recording rate of 25~Hz on a CMOS camera Dalsa Falcon 1.4M100. The field of observation covers an area of  $215~\mu\mathrm{m} \times160~\mu$m.  

\subsection{HT-TPM procedure of analysis}
\label{analysis}

The software developed by Magellium (France) tracks in real time the positions of all the particles, corrects for experimental drift, calculates the asymmetry factor to select tethered particles valid for the analysis, and finally the experimental root mean square end-to-end distance projected on the surface, $R_{\rm exp ||}\equiv \sqrt{\langle \mathbf{R}_{||}^2\rangle}$ the amplitude of motion of the bead, along the time trace (see Fig.~\ref{TPM}). We invite the reader to refer to Ref.~\cite{Manghi2010} for the detailed calculations of $R_{\rm exp ||}$.

In order to measure the small differences expected on $R_{\rm exp ||}$, we set up a two-step procedure for the analysis of $R_{\rm exp ||}$ where, first, a criteria of validity, then some corrections are applied. All this procedure was performed with homebuilt Mathematica scripts. Details can be found in Ref.~\cite{Brunet2015}. During this procedure, around 12\% were eliminated and the final number of kept trajectories for each DNA condition typically ranges between 100 and 1000 (see Tables~1 and~2 of the Supporting Information).

\section{Experimental results}
\label{expresultsect}

\subsection{Simple estimate of DNA persistence length from the amplitude of motion}
\begin{figure}[!t]
(a)\includegraphics[width=\columnwidth]{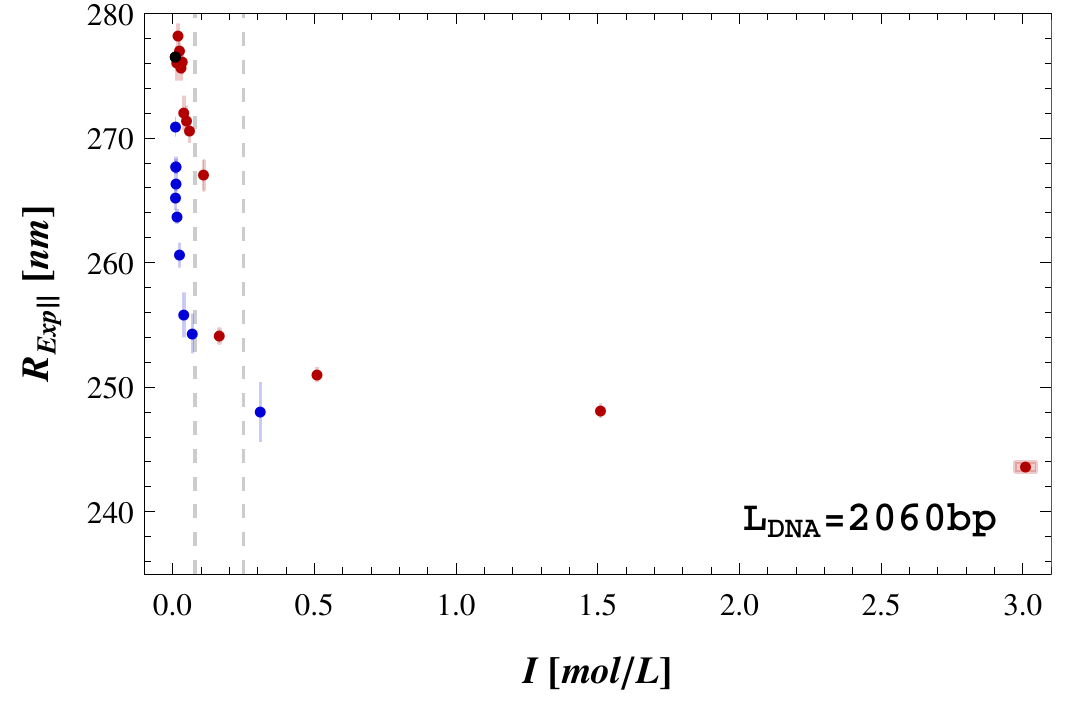}\\
(b)\includegraphics[width=\columnwidth]{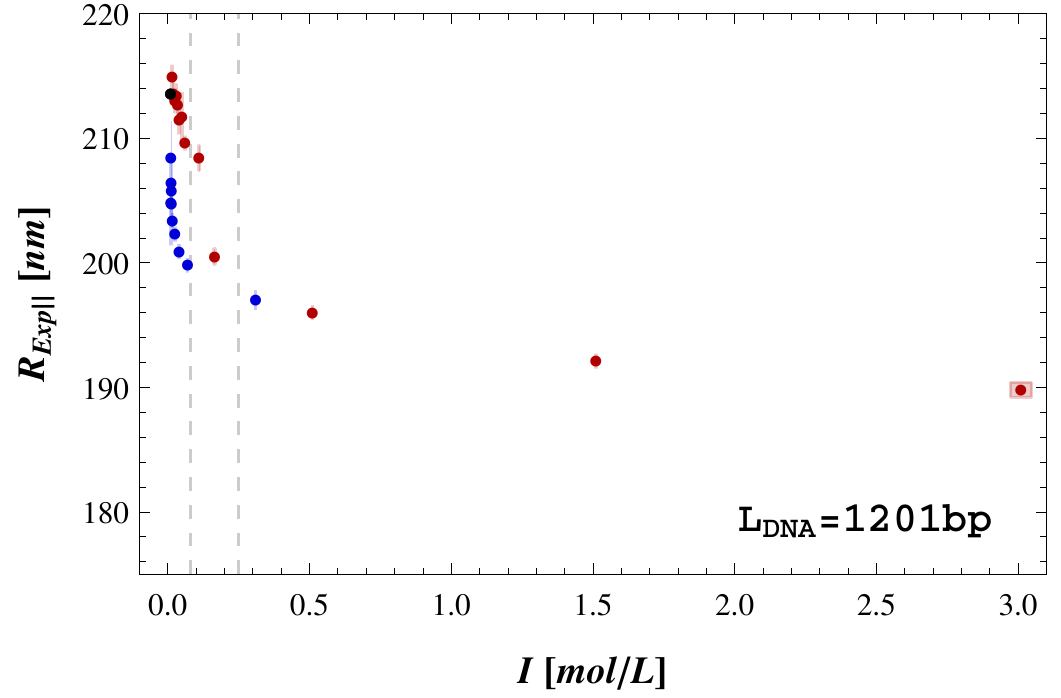}
\caption{\label{figRexp} Experimental HT-TPM amplitude of motion vs. the ionic strength $I$ for a DNA of length (a)~$L=2060$~bp and (b)~$L=1201$~bp with Na$^+$ (red) and with added Mg$^{2+}$ (blue) cations. Vertical lines delimit the close-to-physiological salt conditions, as listed in Table~\ref{LpPhysio}.}
\end{figure}

In Figures~\ref{figRexp}a and~b are shown the variations of the experimental amplitude of motion, i.e. the root mean square end-to-end distance projected on the surface, $R_{\rm exp ||}$, measured by HT-TPM, as a function of the ionic strength, $I$, of the various buffers listed in Tables~1 and~2 (Supporting Information) for two DNA of lengths 2060 and 1201~bp. The ionic strength is defined as 
\be
I=\frac12\sum_i z_i^2 c_i
\ee
where $z_i$ and $c_i$ are respectively the valency and the concentration of ion $i$ in the buffer. The red (resp. blue) symbols correspond to the monovalent Na$^+$ ions (resp. with divalent Mg$^{2+}$ ions added) buffers. The black ones correspond to Zero-salt-buffer conditions.

The plots clearly show that, when $I$ is raised, $R_{\rm exp ||}$ decreases of about 10\%, for monovalent as well as divalent ions. This is thus a subtle effect. This behavior is qualitatively due to the fact that negatively charges carried all along the DNA, repel themselves more strongly, when the ionic strength is low, this repulsion decreasing when the ionic strength increases due to electrostatic screening. This self-repulsion increases the rigidity of the macromolecule thus increasing $R_{\rm exp ||}$. Furthermore, one notices that the variation of $R_{\rm exp ||}$ is amplified when the DNA length, $L$, is increased. This is consistent with the fact that the process under examination occurs homogeneously along the DNA. 

Comparing the experimental results for monovalent ions and divalent ones on the same $I$ range, one observes that the decrease of $R_{\rm exp ||}(I)$ is more important with Mg$^{2+}$ ions, once again an effect amplified for long DNAs.
Moreover, one notes a sudden drop of $R_{\rm exp ||}$ at very low concentration of Mg$^{2+}$ (from $I=10.1$~mmol/L, without  Mg$^{2+}$, to $10.5$~mmol/L). However, it is difficult to know with high precision the ionic strength for such low values, since we cannot exclude that some additional ions get released from the surface, for instance. Therefore, the errors bars are potentially large for these points.

To access to properties of the DNA itself, it is necessary to correct the effect of the particle and the glass substrate to access to properties of DNA only, such as its end-to-end distance, $R_{\rm DNA}\equiv\sqrt{\langle {\bf R}^2_{\rm DNA}\rangle}$. The simplest way is to substract the particle radius by assuming that the DNA extremity and the particle, of radius $R_{\rm p}=150$~nm, move independently, which leads to~\cite{Brunet2015}
\be
\langle {\bf R}^2_{\rm DNA}\rangle = \frac32 \langle  {\bf R}^2_{||}\rangle-R^2_{\rm p}
\label{basic}
\ee
By doing so, we neglect the effect of the glass substrate. It has been taken into account analytically only for long and flexible polymers~\cite{Segall2006}, which is not the case for the DNA of this study. Indeed their length ($L=2060$ and 1201~bp, i.e. 700~nm and 408~nm, using 1~bp$=0.34$~nm) is no more than a few persistence lengths ($L_p\approx50$~nm), which allows us to qualify them as semi-flexible. As already  shown in~\cite{Brunet2015} the approach proposed in~\cite{Segall2006} does not work well for such DNA lengths~\footnote{The approach proposed in~\cite{Segall2006} yields essentially the same result as \eq{basic} for $N=2060$ and is not valid for $N=1201$.}.

In Figures~1a and~b of the Supporting Information are shown the DNA end-to-end distance, $R_{\rm DNA}$, obtained by~\eq{basic}, as a function of the inverse of the ionic strength, $I^{-1}$. Of course the relative effect of the salt is slightly higher, once the particle radius is deduced.
The next step consists in extracting the DNA persistence length $L_p$ from $R_{\rm DNA}$. The simplest way would be to use the Worm-Like Chain (WLC) formula, valid for a phantom chain in solution, resulting from \eq{WLCcorr}~\cite{DoiEdwards}:
\be
\langle {\bf R}^2_{\rm DNA}\rangle  = 2L_p^2\left(\frac{L}{L_p}-1+e^{-L/L_p}\right)
\label{WLC}
\ee
However, this way of extracting $L_p$ leads to quite high values of the persistence length compared to the commonly accepted values around $50$~nm. For instance at low $I$, $L_p$ saturates around 76~nm for the 2060 bp long DNA and 68~nm for the 1201 bp one (data not shown). These high values might be due to the particle--substrate interaction, or particle--polymer or monomer--monomer excluded volume interactions which may swell the DNA. 

\subsection{Refined extraction of the DNA persistence length using simulations of the HT-TPM setup}

To check these effects and the approximations used in \eqs{basic}{WLC}, we performed numerical simulations.
DNA-particle conformations are computed numerically by exact sampling~\cite{Segall2006}. The labeled DNA polymer is generated as a random walk of $N$ steps, corresponding to the links of length $2a$, with a bending energy by step, $E_{\rm bend}=-\kappa_b \cos \theta$, where $\kappa_b$ is the bending modulus and $\theta$ is the angle between successive steps. The number of steps $N$ was chosen such that $a=6$~bp, corresponding to the DNA diameter. The starting point is on the substrate and at each step self-intersecting trajectories (respectively trajectories intersecting the substrate) are eliminated to take into account intra-chain excluded volume interactions (resp. repulsive interaction with the substrate). Hence the polymer is modeled by a chain of beads of (excluded) volume $v=4\pi a^3/3 \simeq36$~nm$^3$, taken constant as a function of the ionic strength (see Fig.~\ref{TPM}). The salt effects are therefore supposed to be completely taken into account in the bending modulus, $\kappa_b$. The last step of length $a+R_p$ has a uniformly random orientation. The persistence length value is related to $\kappa_b$ by $L_p = 2a \beta\kappa_b$, where $\beta=(k_BT)^{-1}$ is the inverse of the thermal energy~\footnote{Strictly speaking, the discrete WLC persistence length leads to $L_p=-2a/\ln[\coth(\beta\kappa_b)-1/(\beta\kappa_b)]$. For $L_p\in[35,70]$~nm, the error is less than 0.25~nm when using the approximation $L_p\simeq 2a \beta\kappa_b$.}.
The two-dimensional projection of the particle position, ${\bf R}_{||}$, was measured and the amplitude of motion, defined as $R_{||}\equiv \sqrt{\langle{\bf R}^2_{||}\rangle}$, is averaged over several millions independent trajectories. 
Since at this level of coarse-graining, electrostatic interactions are not included, we varied $\kappa_b$ by hand such that $L_p$ spans the range 36 to 70~nm, in order to reflect the stiffening due to the decrease of the ionic strength in the solution.
\begin{figure}[!t]
(a)\includegraphics[width=\columnwidth]{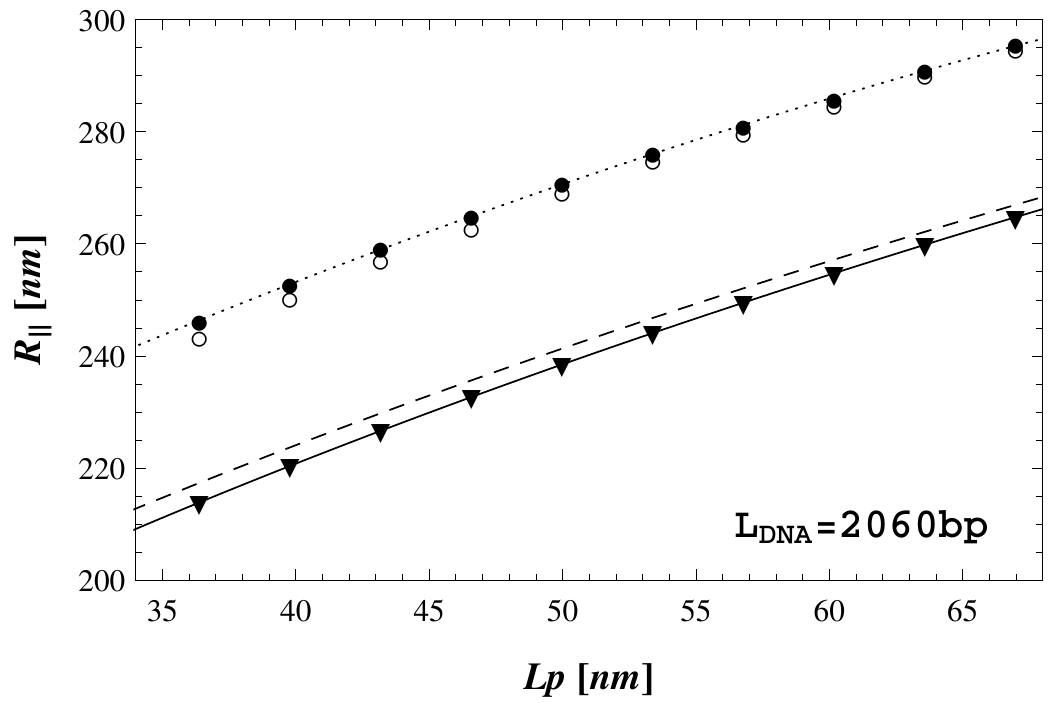}\\
(b)\includegraphics[width=\columnwidth]{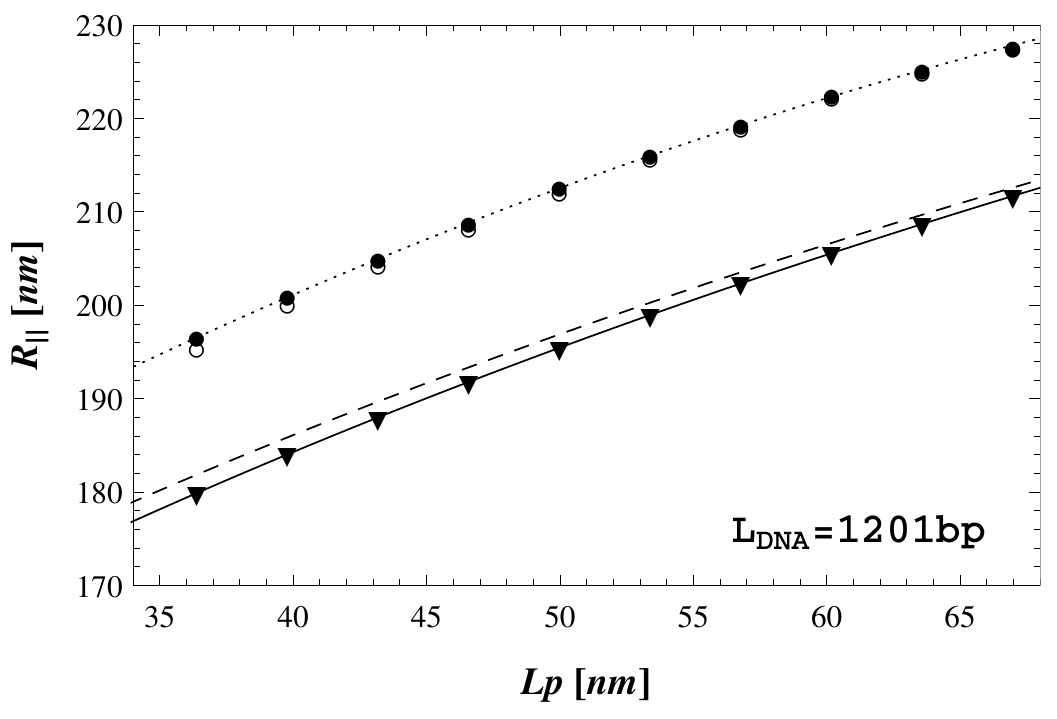}
\caption{Simulated $R_{||}$ vs. $L_p$ for DNA of (a)~2060~bp ($N=172$) and (b)~1201~bp ($N=100$). Triangles correspond to a free polymer-particle complex without excluded volume, full (resp. open) circles to a grafted polymer in the TPM geometry with a hard core particle and with (resp. without) intra-chain excluded volume interactions (see text). The solid line is the discrete WLC formula, \eq{DWLC}, the dashed one is the continuous WLC one, \eq{WLC}. The dotted lines are fits by polynomials: (a)~$R_{||}=-10.7\times10^{-3} L_p^2+ 2.71 \,L_p + 162.1$~[nm], (b)~$R_{||}=-9.0\times10^{-3} L_p^2+ 1.955 \,L_p + 137.35$~[nm]. \label{fig_simu}}
\end{figure}

Figure~\ref{fig_simu} shows the simulated $R_{||}$ as a function of $L_p$ (full circles). We observe an increase of $R_{||}$ from 240 to 300~nm for the 2060~bp long DNA, and from 195 to 230~nm for the 1201~bp one. Both ranges of $R_{II}$ values contain the corresponding experimental observations, which thus indicates that we explored the good range of persistence length values.

For the purpose of comparison are also plotted the end-to-end distance of a free polymer--particle complex without any excluded volume interactions and without wall (triangles), and the polymer--particle plus substrate without excluded volume (open circles).
The solid curves correspond to the discrete WLC result for the end-to-end distance, using \eq{basic} (to include the particle contribution) and without any excluded volume and wall~\cite{Palmeri2008}:
\be
\langle  {\bf R}^2_{||} \rangle = \frac23\left[a^2 N W_N(v(\kappa_b)) + R_{\rm p}^2\right]
\label{DWLC}
\ee
where
\be
W_N(x) = \frac{1+x}{1-x}-\frac{2x}N\frac{1-x^N}{(1-x)^2}
\ee
and $v(\kappa_b)= \coth(\kappa_b)-1/\kappa_b$. Clearly the solid curve perfectly matches  the simulation results, as expected. The dashed one corresponds to the continuous WLC (no excluded volume and no wall), \eq{WLC}, which gives slightly larger end-to-end distances.

One observes that the presence of the particle which interacts both with the substrate and the chain induces a non-constant shift to higher values of $R_{||}$ (from triangles to full circles). Since the intra-chain excluded volume swells the polymer by less than 2~nm, especially for small values of $L_p$ (more flexible chains), the main difference comes from the substrate--particle interactions. This is the reason why the extraction of $L_p$ using \eqs{basic}{WLC} or equivalently \eq{DWLC}  overestimates $L_p$ of about 20~nm.

\begin{figure}[!t]
(a)\includegraphics[width=\columnwidth]{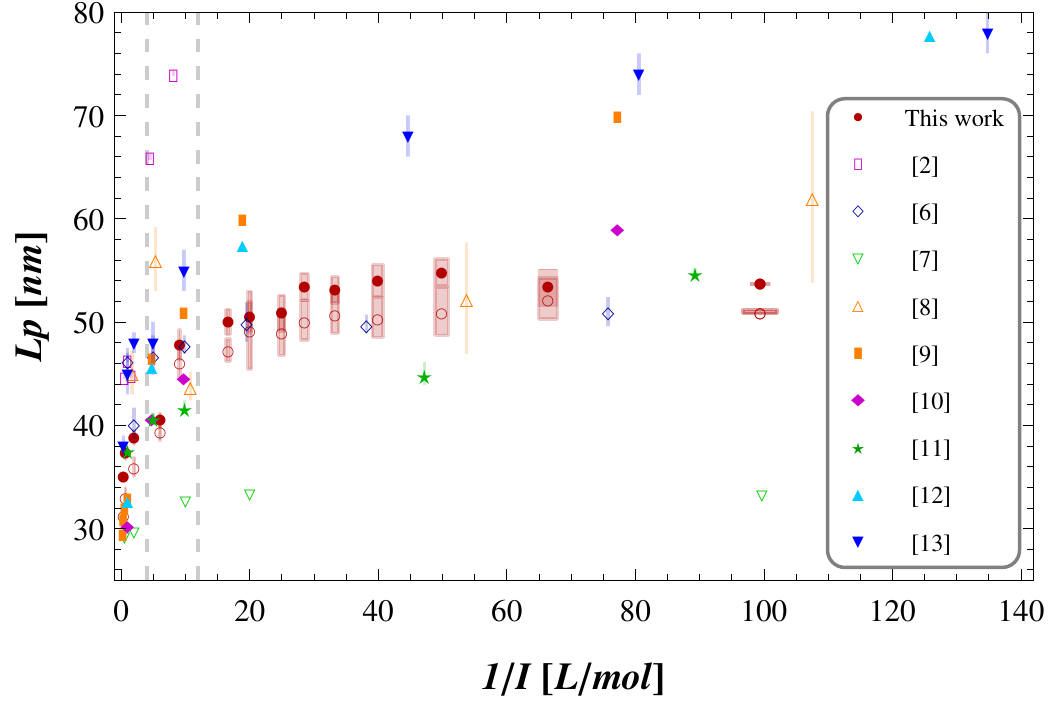}\\
(b)\includegraphics[width=\columnwidth]{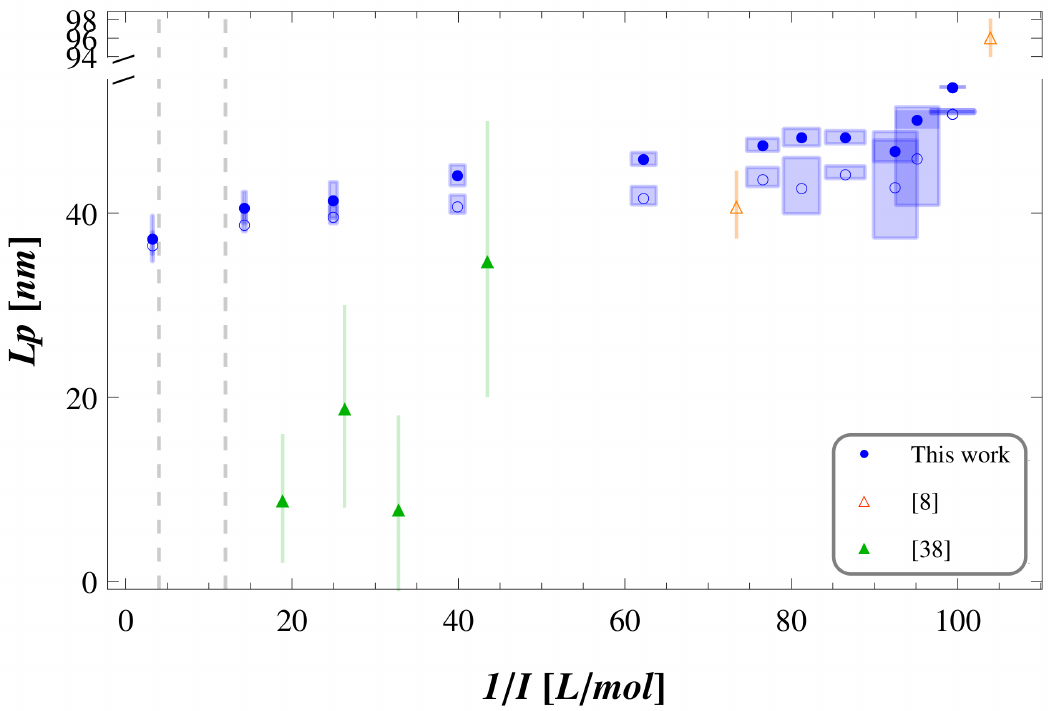}
\caption{DNA persistence length, $L_p$, vs. the inverse of the ionic strength, $I^{-1}$, extracted from the HT-TPM data (full circles for $L=2060$~bp and open ones for $L=1201$~bp), and with available data (Refs. are given in the legend). (a)~Case of buffers with monovalent Na$^+$ salt counter-ions, and (b)~with added divalent Mg$^{2+}$ counter-ions. \label{Lplit}}
\end{figure}

To obtain precise values of $L_p$ from the experiments, we thus fitted the simulation data $R_{||}(L_p)$ by a quadratic polynomial law (see Fig.~\ref{fig_simu}), which in turn allows us to accurately determine the experimental $L_p$ from the experimental values of $R_{||}$. 
The persistence length is then plotted as a function of $I^{-1}$ in Fig.~\ref{Lplit} for the two DNA lengths and the two types of counter-ions (figures are given in Tables~1 and~2 of the Supporting Information). Other available data found in the literature are also shown.

With Na$^+$ counter-ions, $L_p$ values are in the same range both for $L=2060$~bp and 1201~bp, which tends to confirm that the persistence length extracted with this procedure is almost independent of the DNA length, as expected. It increases monotonically from roughly 35~nm for high ionic strength ($I\simeq3$~mol/L) to 54~nm for low one ($I=10$~mmol/L), which corresponds to an increase of more than 50\%. Near physiological salt conditions, around 150~mmol/L, we find $L_p\simeq43$~nm. Moreover using this plot representation, the data show a clear concave shape.

With Mg$^{2+}$ counter-ions, $L_p$ is greatly reduced which is a signature of the role of the ion valency $z$. This has already been observed in previous experiments~\cite{Baumann1997,Hagerman1981}. The $L_p$ values, between 35 and 50~nm, are slightly different for the two DNAs, $L_p$ being larger by almost 5~nm at low $I$ for the longest DNA. Moreover, we observe an abrupt decrease of $L_p$ between the case of no divalent ions and the previous $I^{-1}$ value, corresponding to the addition of 0.15 mmol/L of Mg$^{2+}$. At higher $I$, the increase of $L_p$ is almost linear in $I^{-1}$.

In the following we shall try to fit the so-obtained DNA persistence lengths using the available theories found in the literature. Before this, we compare our experimental values to the ones found in the literature.

\section{Comparison with other experimental data}
\label{comp_exp}

\subsection{Influence of Na$^+$ monovalent ions}

We start the comparison with other experimental studies by focusing on the measurements of $L_p$ for values of $I$ close to the physiological salt conditions.The behaviour of $L_p$ under the action of ionic strength is then discussed.

\subsubsection{Persistence length close to the physiological salt concentration}

An inventory of the values of $L_p$  measured near the physiological salt conditions, \textit{i.e.} for $I \in [100,200]$~mmol/L with only monovalent Na$^+$ counter-ions, is presented in Table~\ref{LpPhysio}. The mean value of these $L_p$ values found in Refs.~\cite{Baumann1997, Sobel1991, Manning1981, Post1983, Kam1981, Rizzo1981, Cairney1982, Porschke1991} is $48 \pm 6$~nm which is in good agreement with our interpolated value $L_p(150~\mathrm{mmol/L}) =43 \pm 3$~nm. 
In fact, the measured $L_p$ varies widely with the experimental techniques in use, the studied DNA which lengths varies from 6646~bp to 50~kbp, as well as with the theoretical and analytical tools used to extract $L_p$. In addition, the variability of these $L_p$ values might come from the  difficulty to perfectly control the presence of divalent ions such as Mg$^{2+}$, the presence of which can have a dramatic effect even at low concentrations (on the order of mmol/L), as mentioned before. 
Finally, it can be noted that the commonly accepted value of $L_p=50$~nm for a ``random''  DNA sequence at physiological salt conditions, \textit{i.e.} $I\simeq 150$~mmol/L ($I^{-1}\simeq 6.7$~L/mol) with only monovalent Na$^+$ counter-ions~\cite{Hagerman1988} slightly exceeds the experimentally derived ones.

\begin{table}[t]
\begin{tabular}{| c | c | c | c | c |}
\hline
$I$ & $L_p$  & $L_{\rm DNA}$ & Experimental & Ref. \\
$[$mmol/L]& [nm] & [bp] & method & \\
\hline
\hline
223.0&$66\pm3$&6646&FB ($25^\circ$C)&\cite{Harrington1978}\\
210.8&46.5&6646&DLS ($20^\circ$C)&\cite{Manning1981}\\
210.8&54.7&6646&DLS ($20^\circ$C)&\cite{Post1983}\\
210.8&40.6&6646&DLS ($20^\circ$C)&\cite{Kam1981}\\
204.1&$48\pm2$&39936&DLS ($20^\circ$C)&\cite{Sobel1991}\\
201.0&$40.6 \pm0.4$&39936 & FB ($25^\circ$C)&\cite{Cairney1982}\\
201.0&$46.8\pm0.4$&39936 & LD&\cite{Rizzo1981}\\
186.2&$56\pm3$&48502 &FOT ($25^\circ$C)&\cite{Baumann1997}\\
165.1&$40.5 \pm0.4$&2060&HT-TPM ($25^\circ$C)&this work\\
165.1&$39.5 \pm0.5$&1201&HT-TPM ($25^\circ$C)&this work\\
154.0&$50\pm5$&434, 587&TEB ($20^\circ$C)&\cite{Hagerman1981}\\
123.0&$74\pm3$&6646&FB ($25^\circ$C)&\cite{Harrington1978}\\
110.1&$47.8 \pm0.7$&2060&HT-TPM ($25^\circ$C)&this work\\
110.1&$46.2\pm0.8$&1201&HT-TPM ($25^\circ$C)&this work\\
103.1&44.6&6646&DLS ($20^\circ$C)&\cite{Manning1981}\\
103.1&44.6&6646&DLS ($20^\circ$C)&\cite{Kam1981}\\
103.1&$53\pm2$&39936&DLS ($20^\circ$C)&\cite{Sobel1991}\\
102.4&$55\pm2$&39936&DLS ($20^\circ$C)&\cite{Sobel1991}\\
101.1&$43\pm1$&$43-179$&TED ($20^\circ$C)&\cite{Porschke1991}\\
101.0&$47.8 \pm0.4$&39936& LD&\cite{Rizzo1981}\\
101.0&$41.8 \pm0.6$&39936 & FB ($25^\circ$C)&\cite{Cairney1982}\\
93.4&$43 \pm 1$&48502 &FOT ($25^\circ$C)&\cite{Baumann1997}\\
\hline
\end{tabular}
\caption{Summary of $L_p$ measured in the close-to-physiological salt conditions (with Na$^+$) found in the literature. In order to compare the whole set of published data~\cite{Baumann1997,Sobel1991,Manning1981,Post1983,Kam1981,Rizzo1981,Cairney1982,Porschke1991}, we rigorously computed the ionic strength used in these data by directly taking experimental values when available, or the values deduced by interpolation otherwise. \label{LpPhysio}}
\end{table}

\subsubsection{Variation of $L_p$ on the whole $I$ range}

In order to compare the global behavior of $L_p(I)$ that we measured with the previously published results, we superimposed all the results in Fig.~\ref{Lplit}a where $L_p$ values are plotted as a function of $I^{-1}$. As mentioned in the Introduction, Savelyev~\cite{Savelyev2012} has recently reviewed all the available experimental data and showed that they could be divided into two groups based on the difference in $L_p$ behaviours observed at high ionic strength, $0.11\leq I\leq 3$~mol/L (see Fig.~\ref{Lplit}). We will keep this division to compare our results to those obtained by the first group of experimental data~\cite{Harrington1978,Rizzo1981,Maret1983,Baumann1997} that shows a slow decrease of $L_p$ with increasing $I$, then to those obtained by the second one~\cite{Kam1981,Manning1981,Borochov1981,Post1983,Sobel1991,Cairney1982}  that found a significant one.

In the first set of experiments, the authors observed a saturation of $L_p$ at high $I$. More precisely for $I$ exceeding 150~mmol/L, $L_p$ decreased by only 10$\%$ only. Moreover, at low $I$, these publications show an increase in $L_p$ of about 10\% similarly as we do observe on our HT-TPM measurements. It can be noted that Harrington~\cite{Harrington1978} found a larger increase of $L_p$ in the low $I$ range than all the other results.

This first group of experimental works gathers results obtained by FB of Harrington~\cite{Harrington1978}, MB of Maret \textit{et al.}~\cite{Maret1983}\footnote{We re-calculated $L_p$ values (between 30 and 90~nm) of Maret \textit{et al.} by using the raw data collected in Table~1 and Eq.~(5) of~\cite{Maret1983}. Our calculation does not correspond to the reported data by Savelyev \textit{et al.}~\cite{Savelyev2010, Savelyev2012}.}, FOT of Baumann \textit{et al.}~\cite{Baumann1997}, and LD of Rizzo \textit{et al.}~\cite{Rizzo1981}. 
The FB, MB or LD methods may be prone to perturbative Joule heating and bulk electrophoresis effect; to minimize them, the ionic strength was therefore kept low.  As FOT method induces a perturbation of the sample structure in the high force regime, we only consider the $L_p$ values obtained in the low force regime using the inextensible WLC model as a comparison to our HT-TPM results, where no force is applied. It is important to note that these studies were performed on DNA exceeding 40 kbp long, and that only scarce measurements were performed at high ionic strength, in opposition to the second group of data.

In this regime of high ionic strength, the second group did not show any plateau but rather a significant decrease of $L_p$ of  about 25 to 30\%. This observation is in good agreement with our measured decrease of 25\% in the same $I$ range. In the low ionic strength range, $0.01\leq I \leq 0.1$~L/mol, this second group measures a but regular decrease of $L_p$ of about 15\% when $I$ increases which is larger than the one we measured, about 10\%. 
FB data~\cite{Cairney1982} can be classified in this set of data, given the 25$\%$ increase of $L_p$ in the high $I$ range. However, variations in the low ionic strength range seems to be modest in comparison to the other publications of this set but closer to ours.

This second group of experimental data is essentially based on the DLS technique performed by Sobel \textit{et al.}~\cite{Sobel1991} and Kam \textit{et al.} (1982)~\cite{Kam1981}. Manning~\cite{Manning1981} and Post~\cite{Post1983} proposed corrections to the extracted $L_p$ values from the original data of Borochov \textit{et al.}~\cite{Borochov1981}. 
In these DLS experiments, the DNA  hydrodynamic radius is deduced from the diffusion coefficient measurement. To infer $L_p$, the usual Gaussian polymer model was used. The number of Kuhn segments being large for the 6646~bp (resp. $\sim$ 40~kbp) DNA under study, $N\simeq 22$ (resp. $N\simeq 1333$), the swelling of the chain was induced as a result of excluded volume. Therefore, estimating precisely the excluded volume is essential. It is nevertheless a challenging task. For instance, Manning's~\cite{Manning1981} and Post's~\cite{Post1983} correction led to $L_p$ values differing by 4 nm at $I=8$~mmol/L, to $\sim 1$~nm at $I=1$~mol/L.

Note that we compare studies performed on long DNA, from 6646~bp to 40~kbp, which are in the flexible regime and thus much more sensitive to excluded volume effects than our HT-TPM experiments made on DNA of lengths $L= 2060$ or 1201~bp, which are in the semi-flexible regime.

In this quantitative comparison, we only considered experiments perfomed on the same range of ionic strength induced by Na$^+$ ions. As a result,  Refs.~\cite{Hagerman1981,Hagerman1988,Elias1981MM,Elias1981BP,Diekmann1982,Porschke1986,Porschke1991,Bednar1995,Lu2002}  that studied the effect of very low Na$^+$-induced $I$, as well as Refs.~\cite{Borochov1984,Wang1997} that focused on the influence of  K$^+$ or Li$^+$ ions, or Ref.~\cite{Mantelli2011} that studied the combined effect of multivalent and monovalent ions, were discarded from our comparison. Note that single molecule study in Ref.~\cite{Smith1992} was not considered either due to the too few values of $I$ explored.

\subsection{Influence of Mg$^{2+}$ divalent ions}

The decrease of  $L_p$, that is measured when $I$ varies from 10 to 310~mmol/L, is about 35\% whatever the ion valency and the DNA length (Fig.~\ref{Lplit}). Yet the shape of this decrease as a function of $I$ is however completely different depending on the valency of the ions used. Divalent ions appears to induce an almost linear decrease of $L_p$ as a function of $I^{-1}$, when monovalent ions cause a decrease with a concave shape. In addition, the absolute value for $L_p$ is smaller by about 5~nm with divalent ions than with monovalent ions at low $I$. This observed trend is in agreement with the first quantitative observations at very low $I$ by Hagerman~\cite{Hagerman1981}, Elias and Eden~\cite{Elias1981MM}, and then by others~\cite{Maret1983,Baumann1997}.
In addition, Hagerman~\cite{Hagerman1981} observed on 434 and 587 bp DNA by TEB, the same abrupt fall of $L_p$ at very low $I$ with Mg$^{2+}$ ions followed by a slower decrease, as shown in Fig.~\ref{Lp}. The magnitude of this initial decrease was however larger by around 30\% in~\cite{Hagerman1981}, 60\% in~\cite{Baumann1997} whereas~\cite{Maret1983} found a decrease similar to ours.

Some other studies explored the influence of Mg$^{2+}$ on DNA flexibility~\cite{Porschke1986,Porschke1991,Lu2002,Kumar2014,Han2009,Wang1997} and showed a rise in DNA flexibility with the addition of Mg$^{2+}$ counter-ions in solution. Nevertheless, these studies only probed a few values of $I$, and some KCl was added to the solution preventing any quantitative comparison.

Dietrich \textit{et al.}~\cite{Dietrich2009} also used the TPM technique to monitor the effect of Mg$^{2+}$ on a DNA fragment of 4882~bp and observed a large decrease of $R_{\rm exp||}$ in presence of divalent ions. Yet $L_p$ values extracted from these experimental data, much smaller than any other published experimental data, cannot be quantitatively compared to our results due to several errors in the extraction procedure. The persistence length was extracted assuming that the particle excursion was related to $L$ using a simple Hooke law in the Gaussian regime, $\langle {\bf R}^2_{||}\rangle=2L_p L$, thus seemingly forgetting the factor 3/2 due to dimensionality and not subtracting the particle radius [see \eq{basic}], but also ignoring excluded volume effects. Moreover, they presented a very large error bar of 10~nm, due to the small number of trajectories, ranging between 6 and 27.

In the next section, we compare our experimental values of the DNA persistence length to the various theories developed in the literature. 

\section{Comparison between experiments and existing theories}
\label{comp_th}

\subsection{Odijk-Skolnick-Fixman approach at high ionic strength}

Several models have been proposed in the literature to explain the variation of the persistence length $L_p$ of polyelectrolytes with the ionic strength $I$. When electrostatic interactions between mobile ions and the polyelectrolyte are taken into account at the Debye-H\"uckel level (mean-field level and approximation of small values of the electrostatic potential), it has been shown by Odjik~\cite{Odijk1977} and Skolnick and Fixman~\cite{Skolnick1977} (OSF) that, using a perturbative approach around an infinitely stiff rod, the persistence length has two contributions
\be
L_p^{\rm OSF}=L_p^\infty + \frac{\ell_B}{4A^2\kappa^2}
\label{OSF}
\ee
where $L_p^\infty$ is the bare persistence length (in the limit $\kappa A\to\infty$), and the second term is an electrostatic contribution to $L_p$, where $A$ is the distance between elementary charges along the chain and 
\be
\kappa = (8\pi \ell_B I)^{1/2}
\ee
is the Debye-H\"uckel screening parameter. The Bjerrum length, $\ell_B=e^2/(4\pi \epsilon k_BT)$, is equal to 0.715~nm in water at room temperature which yields $\kappa=3.29\sqrt{I}$~nm$^{-1}$ (where $I$ is in mol/L). \eq{OSF} is theoretically valid for polymer conformations close to the rod-like one, i.e. for $\ell_BL_p^\infty\gg A^2$.

For a dsDNA, with two phosphate anions per bp, we have $A=0.17$~nm, and $L_p^\infty\approx 50$~nm so that the validity of \eq{OSF} is well verified. As a function of the ionic strength $I$, \eq{OSF} can be rewritten as
\be
L_p^{\rm OSF}=L_p^\infty + \frac{0.559}{I}\quad [\mathrm{nm}]
\label{OSF2}
\ee
where $I$ is in mol/L. Note that the numerical value of the coefficient depends on the model as shown by Fixman~\cite{Fixman2010}. In any case, Fixman affirms that the exponent in $I^{-1}$ is robust at large $I$ (large $\kappa$).
Clearly the data shown in Fig.~\ref{LpOSFM}, where the experimental $L_p$ is plotted vs $I^{-1}$, show a concave shape for high ionic strength, and therefore are not well fitted by the linear law, \eq{OSF2}. In Fig.~\ref{LpOSFM} is shown a linear fit (dashed lines), $L_p^\infty +C/I$, for $0<I^{-1}<10$~L/mol, which yields $C=1.21$~nm$\cdot$mol/L and $L_p^\infty=35.5$~nm for $L=2060$~bp, and $C=1.52$~nm$\cdot$mol/L and $L_p^\infty=31.7$~nm for $L=1201$~bp, which are almost 3 times larger than the one predicted by OSF. Since we do not have enough data at very high $I$ with added Mg$^{2+}$, we did not try to fit \eq{OSF2} for the Mg$^{2+}$ case.

The data obtained by the first group~\cite{Harrington1978,Rizzo1981,Maret1983,Baumann1997} at high $I$ were qualitatively in agreement with the OSF theory, in the sense that the variations of $L_p$ are small at high $I$. However, no fit using \eq{OSF} was done in these works.

\subsection{Manning charge renormalization at low ionic strength}

The OSF theory fails to explain the variation of $L_p(I^{-1})$ for the whole range of $I$ studied. Indeed it does not reproduce the concave shape observed in  Fig.~\ref{Lp}. A plausible explanation is that the Debye-H\"uckel approximation of small values of the electrostatic potential is not valid for double-stranded (ds)DNA, which is highly charged. Indeed, Manning has shown in the 60's that, in the limit of low $I$, some counter-ions somehow ``condense'' close to the DNA due the large net charge of DNA~\cite{Manning1969}. This effect is a non-linear effect associated with the mean-field Poisson-Boltzmann equation close to charged cylinders. This phenomenon, known as the Manning condensation, tends to reduce the effective charge of the DNA. Manning proposed that, in the \textit{low salt limit}, the effective linear charge density is $\alpha/A$ where $\alpha$ depends on the parameter 
\be
u=\frac{z\ell_B}A
\ee 
where $z$ is the counterion valency: If $u<1$, $\alpha=1$ and if $u>1$ then $\alpha=1/u$. For dsDNA, the Manning parameter is $u=4.21$ for Na$^{+}$ counter-ions ($z=1$) and $u=8.41$ for Mg$^{2+}$ ones ($z=2$). Thus for monovalent counter-ions, $\alpha=0.24$ and the effective charge is decreased by a factor of roughly 75\%. For a mixture of counter-ions with different valencies, which is the case in our experiments with Mg$^{2+}$ ions, the effective charge is $\alpha=A/(z\ell_B)$ where $z$ is the largest valency, i.e. only counter-ions with the largest valency (here divalent ones) condense along the DNA and no monovalent ions are condensed (unless divalent counter-ions are depleted, which is not the case in most of the experiments and especially in ours)~\cite{Manning1984}.

The OSF equation, \eq{OSF}, has thus been modified according to~\cite{LeBret1982,Fixman1982}
\be
L_p^{\rm OSFM}=L_p^0 + \frac1{4z^2\ell_B\kappa^2}
\label{OSFMan}
\ee
which changes the slope $B$ of the linear relationship $L_p=L_p^0+B/I$ from 0.559 [\eq{OSF2}] to 0.033~nm$\cdot$mol/L, and should be valid at low ionic strength, i.e. for large $I^{-1}$. This is the reason why the constant $L_p^0$ is \textit{a priori} different from the one of \eq{OSF}, $L_p^\infty$. Note that adopting the Manning condensation, valid at low $I$, to the OSF calculation of the persistence length, valid at high $I$ is somewhat not consistent. Moreover \eq{OSFMan} is not a proper asymptotic expansion in the limit $\kappa A\to 0$ since it leads to a diverging $L_p$. Of course, this limit cannot be reached in practice, since DNA counter-ions and ions resulting from water dissociation are always present even when no salt is added, which ensures that $I\neq0$ (even if it can be very small). Theoretically, the salt-free case corresponds to $\kappa^{-1}\gg L$, i.e. to ionic strengths $I \ll 10^{-6}$~mol/L, which is far from being the case in our experiments.
\begin{figure}[!t]
\includegraphics[width=\columnwidth]{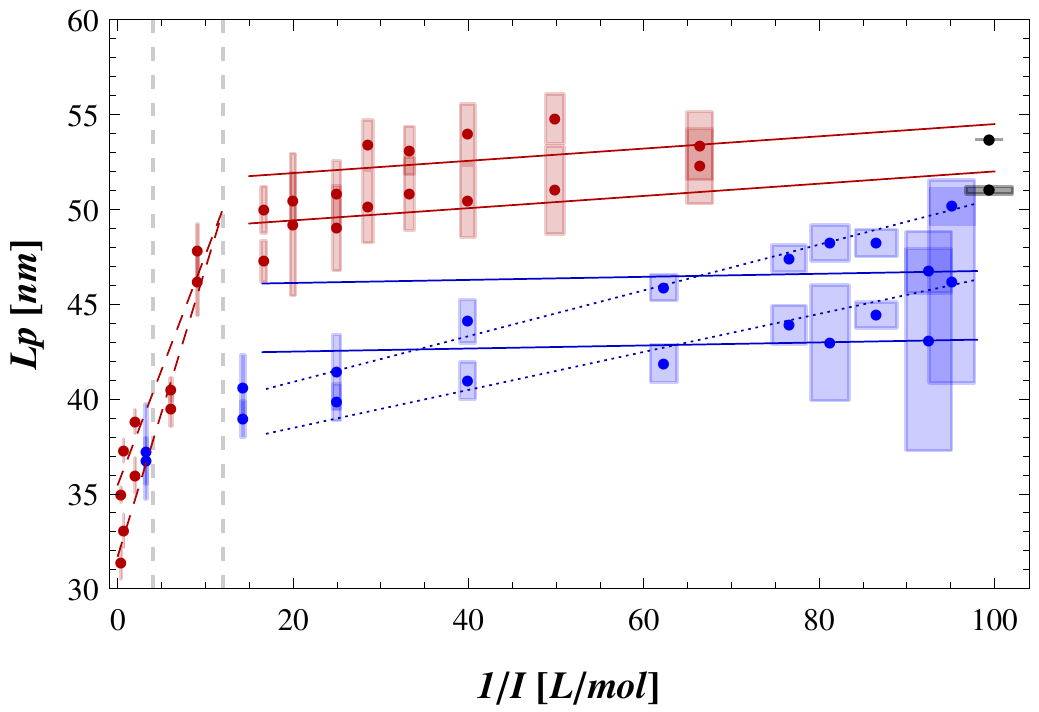}
\caption{Linear fits of the persistence length $L_p$ measured by HT-TPM (already shown in Fig.~\ref{Lplit}) vs. the inverse of the ionic strength $I^{-1}$.  Dashed lines correspond to \eq{OSF} for large $I^{-1}$, $L_p^\infty +C/I$, with 2 fitting parameters (see text for values), and solid lines to \eq{OSFMan} for low $I^{-1}$, with $L_p^0$ as fitting parameter. Better fits are obtained  for Mg$^{2+}$ by leaving $B$ free (dotted lines, $L_p^0=38.5$~nm, $B=0.121$ for $L=2060$~bp, and $L_p^0=36.5$~nm, $B=0.1$ for $L=1201$~bp).\label{LpOSFM}}
\end{figure}

We fitted \eq{OSFMan} on our experimental $L_p$ values at low salt for the Na$^+$ case keeping $L_p^0$ as a free parameter. The results, shown in Fig.~\ref{LpOSFM} (solid lines), are quite satisfactory with $L_p^0=51.3$~nm for $L=2060$~bp and 48.8~nm for $L=1201$~bp (letting $B$ free leads to a slightly higher value of $B=0.038$~nm$\cdot$mol/L). However \eq{OSFMan} does not fit the $L_p$ values for the Mg$^{2+}$ case, the slope being larger (we found $B\simeq0.1$~nm$\cdot$mol/L) whereas \eq{OSFMan} predicts a slope divided by $z^2=4$, $B=0.008$~nm$\cdot$mol/L.

The persistence lengths measured in~\cite{Rizzo1981,Maret1983,Baumann1997,Wenner2002} at low $I$ with Na$^+$ ions were fitted by $L_p=L_p^0+B/I$. In~\cite{Baumann1997} $B=0.033$~nm$\cdot$mol/L was fixed to the Manning value, and $L_p^0$ was found to be 45 to 50~nm, which is comparable to our results.
Maret \textit{et al.}~\cite{Maret1983} fixed $L_p^0=50$~nm and obtained reasonable fits with $0.024<B<0.041$~nm$\cdot$mol/L, which suggests a large error bar on the experimental values. Rizzo \textit{et al.}~\cite{Rizzo1981} also fitted their data for $3<I<1000$~mmol/L. They obtained $L_p^0=46 \pm 1$~nm in agreement with our value (see Fig.~\ref{LpOSFM}), and $B=0.043$~nm$\cdot$mol/L. Wenner \textit{et al.}~\cite{Wenner2002} used measured $L_p$ of ds DNA for various Na$^+$ concentrations by fitting force-extension curves at low forces. They obtained $L_p^0=46$~nm and $B=0.037$~nm$\cdot$mol/L. All these values of $B$ are in agreement with ours and the one predicted by \eq{OSFMan}, 0.033~nm$\cdot$mol/L.

Tomic \textit{et al.}~\cite{Tomic2006} did dielectric spectroscopy experiments on semi-dilute DNA solutions with NaCl to investigate the high frequency and low frequency relaxation modes vs. added salt concentration strength. In the high added salt limit (and relatively low DNA concentration) the length scale of the low frequency relaxation mode, $L_{\rm LF}$, can be interpreted as the DNA persistence length $L_p$. Their results are in qualitative agreement with the OSF--Manning theory, but with a coefficient $B=0.08$~nm$\cdot$mol/L larger than the Manning value, and smaller that the OSF one. Note that this discrepancy can be due to the fact that, in these experiments the total ionic strength is different from the added salt concentration. Using the same experimental method, they investigated the effect of Mg$^{2+}$ in solution~\cite{Grgicin2013}. The $L_{\rm LF}$ was about 1.5 times shorter in Mg-DNA solution than in Na-DNA solution, also suggesting an increased screening with Mg$^{2+}$. The behavior of $L_{\rm LF} $ was again explained by the OSF--Manning theory but with a different value of the effective linear density.

Later, Manning proposed to modify \eq{OSFMan} by multiplying the salt-dependent persistence length by a factor $(2u-1)/u\simeq1.76$ for $z=1$ and 1.88 for $z=2$ which gives a worse result for Na$^+$ and is not sufficient for Mg$^{2+}$ in our case~\cite{Manning2001}. This correction is therefore not suitable.

\subsection{Mean-field non-linear corrections at intermediate ionic strengths}

\begin{figure}[!t]
\includegraphics[width=\columnwidth]{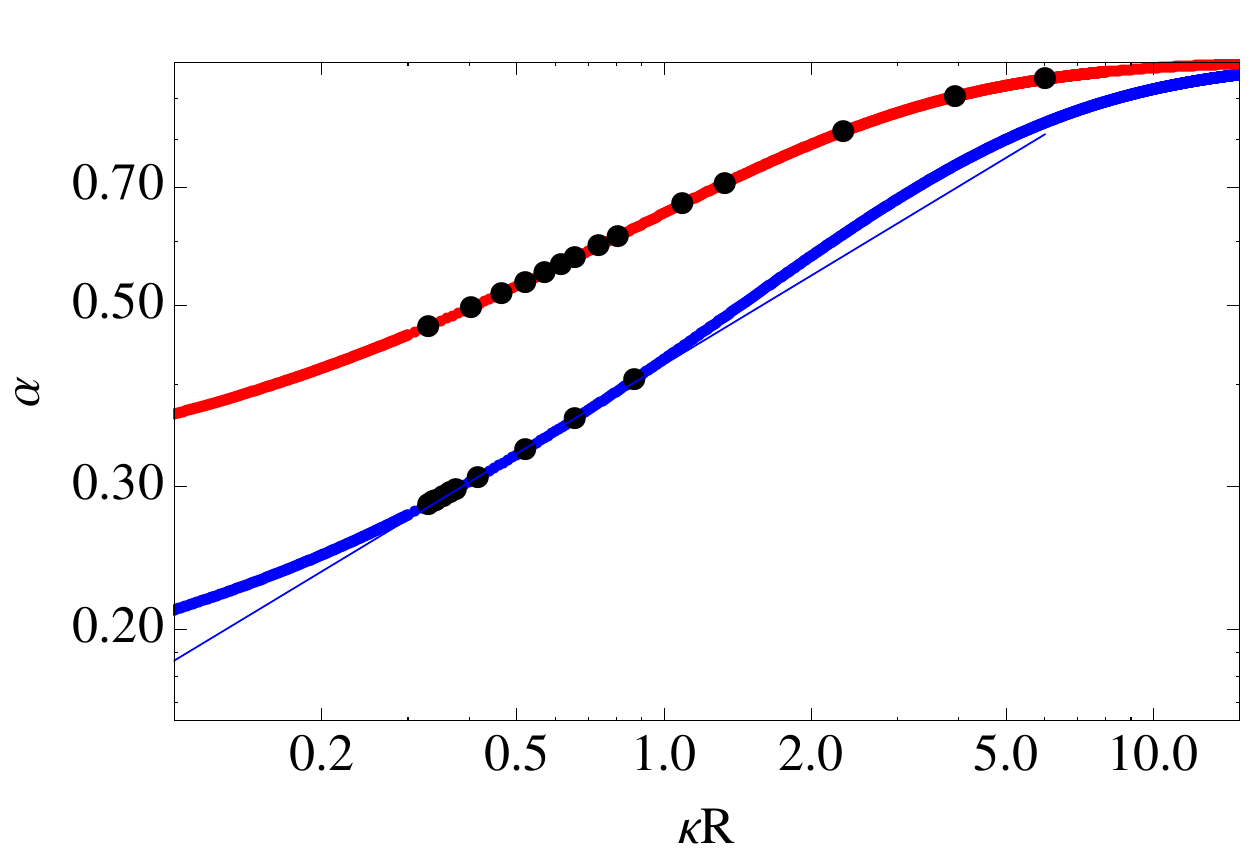}
\caption{\label{fig_alpha} Renormalization charge parameter $\alpha$ vs. dimensionless screening parameter $\kappa R$, solution of \eq{eqVar} for $u=4.11$ (monovalent counter-ions in red) and $u=8.23$ (divalent counter-ions in blue). The black circles correspond to the experimental values studied in Section~III.}
\end{figure}
\begin{figure}[!t]
\includegraphics[width=\columnwidth]{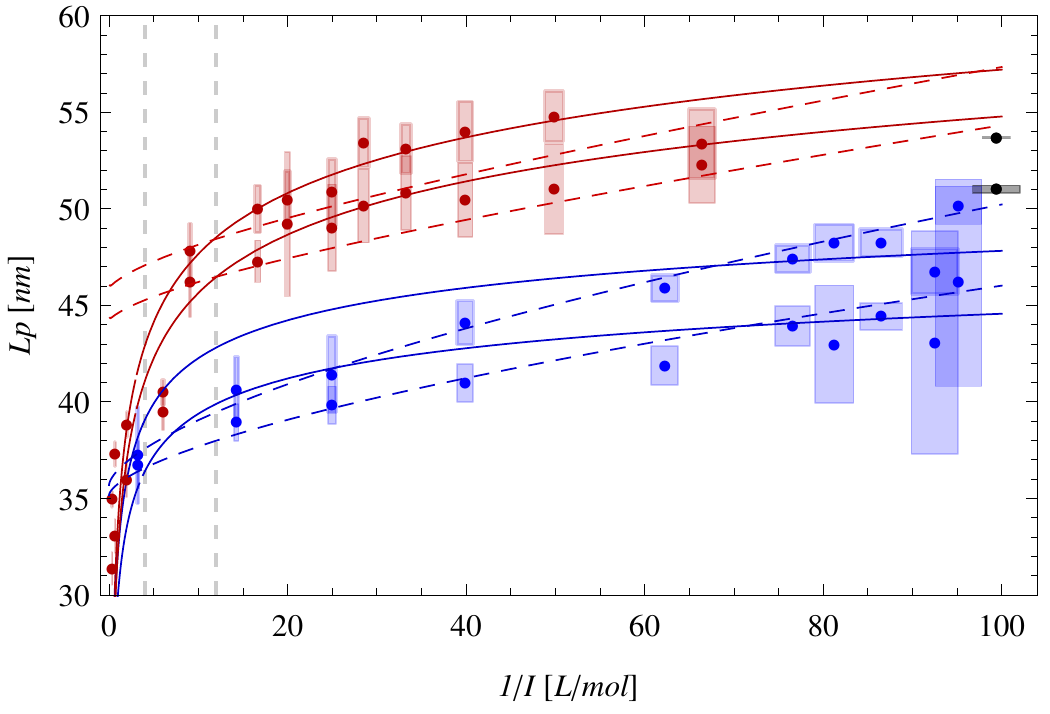}
\caption{Same as Fig.~\ref{LpOSFM} where dashed lines are fits using \eq{Lpvar} and solid lines using \eq{LpManning}. Parameter values are given in the text. \label{Lp}}
\end{figure}

In any case, \eq{OSFMan} does not explain the concave shape shown in Fig.~\ref{Lp} at intermediate ionic strengths. To do so, one must have a renormalization factor of the DNA charge, $\alpha$ that depends on the ionic strength $I$. Such approach has been developed by Netz and Orland~\cite{Netz2003}, in which the Poisson-Boltzmann equation is variationally approximated by a Debye-H\"uckel one with $\alpha$ as a variational parameter which renormalizes the electrostatic potential at the DNA surface. 

Here, we do the calculations by assuming that the DNA is not penetrable to ions, contrary to Ref.~\cite{Netz2003}. The dimensionless electrostatic Debye-H\"uckel  potential, $\phi = ze\beta\psi$, for an electrolyte (valency $z$) of Debye-H\"uckel constant $\kappa$ around a cylinder of radius $R$ and surface charge density $\sigma=(2\pi A R)^{-1}$ empty of ions is:
\bea
\phi (r)=\frac{2u}{\kappa R}\frac{K_0(\kappa r)}{K_1(\kappa R)} \quad \mathrm{for}\quad r>R \label{DHcylinder}\\
\phi (r)=\frac{2u}{\kappa R}\frac{K_0(\kappa R)}{K_1(\kappa R)}\quad \mathrm{for}\quad r\leq R
\eea
where $K_0$ and $K_1$ are the modified Bessel function of order 0 and 1 and $u=z\ell_B/A$ is the Manning parameter.

Following Netz and Orland~\cite{Netz2003}, the full non-linear Poisson-Boltzmann equation is solved variationally by assuming that the solution is of the Debye-H\"uckel form, $\alpha \phi (r)$, where $\phi(r)$ is given in \eq{DHcylinder}, and $\alpha$, the fraction of the ``free" counter-ions, is the variational parameter, and is solution of (see the Supporting Information):
\bea
&&4\pi z\ell_B(1-\alpha)\int d\br \rho(\br) \phi(\br) =\nonumber\\
&&\kappa^2 \int d\br\Omega(\br) \phi(\br)\left\{\sinh[\alpha \phi(\br)]-\alpha\phi(\br)\right\} \label{var}
\eea
where $\rho (\br) =\sigma \delta(r-R)$ is the charge distribution. Note that the ionic exclusion factor $\Omega(\br)$ was incorrectly put just in front of the $\sinh$ in~\cite{Netz2003}. Using \eq{DHcylinder}, \eq{var} simplifies to
\bea
&&2u(1-\alpha) K_0(\kappa R)= \label{eqVar} \\ 
&&\int_{\kappa R}^\infty xK_0(x)\left\{\sinh\left[ \frac{2u\alpha}{\kappa R} \frac{K_0(x)}{ K_1(\kappa R)}\right]- \frac{2u\alpha}{\kappa R} \frac{K_0(x)}{ K_1(\kappa R)} \right\}dx\nonumber
\eea
The solution $\alpha(\kappa R)$, where $R\simeq1$~nm is the DNA radius, is plotted in Fig.~\ref{fig_alpha} for $z=1$ (red) and $z=2$ (blue). The renormalization factor $\alpha$ is a monotonous increasing function of $\kappa R$ with $\alpha(\kappa R\to\infty)\to1$ and $\alpha(\kappa R\to0)\to1/u$. Hence it induces a concave shape to $L_p$ defined as
\be
L_p^{\rm MF}=L_p^0 + \frac{\ell_B}{4A^2\kappa^2}[\alpha(\kappa R)]^2
\label{Lpvar}
\ee
One observes in Fig.~\ref{fig_alpha} that the variations of $\alpha(\kappa R)$ are greater for $z=2$ than for $z=1$. Moreover for $\kappa R\ll1$, $\alpha$ for $z=2$ is smaller than for $z=1$. These two features are in qualitative agreement with what is observed in Fig.~\ref{Lp}.

To do a quantitative comparison of \eq{Lpvar} with the experimental data in the whole range of ionic strengths, we used a polynomial interpolation fonction to fit $L_p$ for monovalent ions ($z=1$), and a power law for divalent ions ($z=2$) $\alpha (\kappa R)\simeq0.423\ \left(\kappa R\right)^{0.364}$, shown in Fig.~\ref{fig_alpha}. The fitting parameters are  $L_p^0$ and a prefactor in front of the second term of the rhs. of \eq{Lpvar} (expected to be close to 1). Fits are shown in Fig.~\ref{Lp} as dashed lines.  Clearly this approach leads to a slightly concave curve for $L_p(I^{-1})$ for the two types of counter-ions. However, whereas fits of Mg$^{2+}$ data (in blue) are reasonably good ($L_p^0=35.7$~nm, prefactor equal to 1.8 for $L=2060$~bp, and $L_p^0=35.2$~nm, prefactor of 1.5 for $L=1201$~bp), the fits of the Na$^+$ data are not good for $I>0.1$~mol/L ($I^{-1}<10$~L/mol), the concavity being not enough pronounced ($L_p^0=44.2$~nm, prefactor of 0.8 for $L=2060$~bp, and $L_p^0=45.9$~nm, prefactor of 0.9 for $L=1201$~bp). Moreover, here again the non electrostatic contribution to the persistence length, $L_p^0$ varies from the fit of Na$^+$ data to the fit of the Mg$^{2+}$ ones.

\subsection{Beyond mean-field: ion-ion correlations and thermal fluctuations}

Other approaches, that go beyond the mean-field approximation by taking into account ion-ion correlations or/and thermal fluctuations, have been proposed~\cite{Nguyen1999,Golestanian1999,Ariel2003EPL}.

Nguyen~\textit{et al.}~\cite{Nguyen1999} considered ion-ion correlation in the strong coupling regime defined as $\Gamma= z^{3/2}\ell_B/\sqrt{2RA}\gg 1$. The theoretical limit $\Gamma\to \infty$ corresponds to the freezing of the strongly coupled counter-ions into a Wigner crystal close to the DNA molecule.  For $\Gamma\gg1$, they found the following correction to the persistence length in the limit of zero ionic strength:
\be
L_p=L_p^0+\frac{R^2}{2zA}(-0.83\,\Gamma+0.33\,\Gamma^{1/4}+0.87)
\label{Nguyen}
\ee
which is independent of the ionic strength $I$. Therefore they propose that $L_p$ is constant at vanishing $I$. For DNA at room temperature, one has $\Gamma= 1.2\,z^{3/2}$, and applying \eq{Nguyen} yields a correction of $+0.65$~nm for $z=1$ (even if $\Gamma\simeq 1$), and $-2.20$~nm for $z=2$. \eq{Nguyen} therefore qualitatively explains the observed abrupt decrease of $L_p$ of about 5~nm, when Mg$^{2+}$ ions are added at very low $I$ in the buffer: the monovalent Na$^+$ ions are replaced by strongly coupled divalent counter-ions, whose correlations decrease the global free energy of the condensed ions and therefore the DNA bending free energy. Hence $L_p(I\to 0)$ decreases with $z$. Following this approach, it is thus consistent to choose two different asymptotic values for $L_p$ when $I\to0$ for monovalent and divalent ions. This explains why these values where slightly different (of about 9~nm) in the preceding section. Note that this theory explains the constant shift at very low $I$ but does not explain the change in the shape of $L_p(I^{-1})$. 

Thermal fluctuations have been taken into account by Golestanian \textit{et al.}~\cite{Golestanian1999}, who obtained a correction to the persistence length due to fluctuation-induced correlations between ions. Indeed, they correct the OSF--Manning formula \eq{OSFMan} at low $\kappa A$, following 
\be
L_p=L_p^0 + \frac{\ell_B}{4u^2(\kappa A)^2} f(\kappa A,u)
\ee 
where $f(\kappa A,u)= [1-2(u-1)\ln(\kappa A)]^{-2}$. Ariel and Andelman~\cite{Ariel2003EPL} proposed a similar correction with $f(\kappa A,u)= u(2-u)-(u-1)^2/[u \ln(\kappa A)]$.

Both formulae do not apply to DNA for $z=1,2$ ($u=4.21$ and 8.41). Golestanian's formula yields a slope which is almost divided by 200 with a convex shape for $I^{-1}< 10$~L/mol, and Ariel's one yields a decreasing function of $L_p(I^{-1})$ as $I^{-1}$ increases for the whole $I$ range. These two theories are therefore not consistent with the whole set of experimental data shown in the experimental results Section. 

\subsection{Manning's internal stretching force calculation for $L_p$}

The major issue in trying to fit the above theories for the whole range of ionic strengths is to find a fit which yields the concave shape observed for the experimental values. \eq{OSF} \textit{a priori} valid for high $I$ and \eq{OSFMan} valid for low $I$ cannot be reconciled since the constant value is clearly different in both cases, $L_p^0>L_p^\infty$ and should therefore vary with $I$.

Manning noted this discrepancy in 2006~\cite{Manning2006} and proposed a new formula for the persistence length by taking into account the internal electrostatic tension due to the repulsion between charges along the polyelectrolyte. He adapted the calculation by Netz~\cite{Netz2001} for strongly stretched polyelectrolytes at the Debye-H\"uckel level to the framework of the counterion condensation approach to obtain the persistence length of a polyelectrolyte as a function of $\kappa A$ and the persistence length of the so-called null isomer (the hypothetical structure of the polyelectrolyte if the backbone charges are set to zero), $L_p^*$:
\bea
L_p &=& \left(\frac\pi2L_p^*\right)^{2/3}\frac{R^{4/3} }{z^2\ell_B}\left[(2u-1)\frac{\kappa A \,e^{-\kappa A}}{1-e^{-\kappa A}}\right. \nonumber\\
&&\left. -1-\ln(1-e^{-\kappa A})\right]
\label{LpManning}
\eea
\eq{LpManning} fits very well our data for the Na$^+$ case on the whole ionic range with only one fitting parameter, $L_p^*$ (Fig.~\ref{Lp}). The fitting values are $L_p^*=6.0$~nm (for $L=2060$~bp) and 5.4~nm (for $L=1201$~bp), close to the value of 7.4~nm fitted by Manning on Baumann's experimental data~\cite{Manning2006}.

Savelyev~\cite{Savelyev2012} performed numerical simulations to investigate the dependence of the persistence length of double-stranded DNA on solution with various ionic strength. A coarse-grained model of two-bead DNA chain with explicit mobile ions (Na$^+$ and Cl$^-$ ions)~\cite{Savelyev2010} was designed to reproduce physical salt conditions from $10^{-4}$ to 0.1~mol/L (the water solvent is implicit). Their numerical results for $L_p(I)$ are in semi-quantitative agreement with \eq{LpManning} for $I>0.1$~mol/L. For lower $I$ the agreement is better with the OSF theory, \eq{OSF} (see Fig.~2 of Ref.~\cite{Savelyev2012}, the fitting parameter values are not given). Moreover, Savelyev~\cite{Savelyev2012} compared previous experimental results to his simulations and found a qualitative agreement.

Assuming that, according to the counter-ion condensation theory~\cite{Manning1984}, all the condensed counterions are divalent, we use the same formula for the persistence lengths with Mg$^{2+}$ counter-ions. It leads to poorer fits (Fig.~\ref{Lp}) with very different values for $L_p^*$: 14.1~nm (for $L=2060$~bp) and 12.7~nm (for $L=1201$~bp). The fact that $L_p^*$ varies, and increases, with the counter-ion valency is puzzling. Hence the dependence of $L_p$ with $z$ in \eq{LpManning} is not consistent with our experimental data.

\section{Interpolation formula for the whole ionic strength range}
\label{interpolation_section}
\begin{figure}[!t]
\includegraphics[width=\columnwidth]{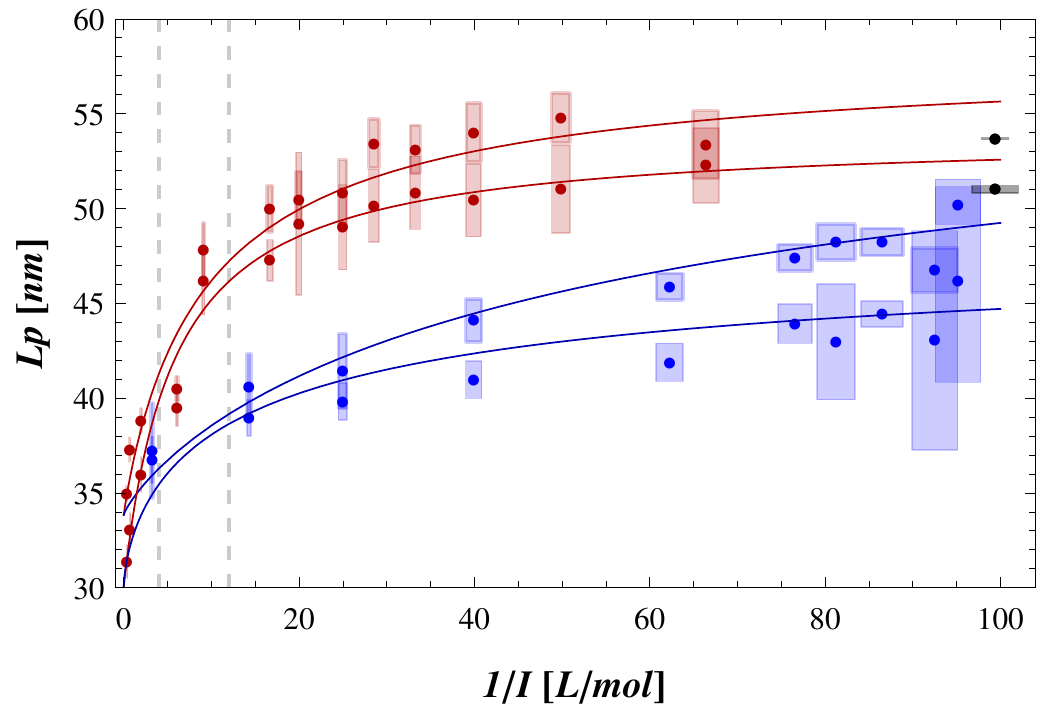}
\caption{DNA persistence length, $L_p$, vs. the inverse of the ionic strength, $I^{-1}$, extracted from the HT-TPM data. The red (resp. blue) symbols correspond to buffers with sodium (resp. magnesium) counter-ions (top curves are for $L=2060$~bp and bottom ones for $L=1201$~bp). The solid lines are fits using \eq{interp} with parameter values are given in the text. \label{Lpinterp}}
\end{figure}

We propose the following interpolation formula to fit the four sets of data (2 DNA lengths, monovalent and divalent salts) on the whole $I$ range:
\be
L_p=L_p^\infty + \frac{L_p^0-L_p^\infty}{1+(I/I_0)^\delta}
\label{interp}
\ee
with 4 fitting parameters, $L_p^0$, $L_p^\infty$, $I_0$, and $\delta$.
The fits, shown in Fig.~\ref{Lpinterp}, are very good for the monovalent Na$^+$ ion, with $L_p^0=58.1$~nm, $L_p^\infty=33.8$~nm, $I_0=0.104$~mol/L, and $\delta=0.931$ for $L=2060$~bp, and $L_p^0=53.9$~nm, $L_p^\infty=30.1$~nm, $I_0=0.174$~mol/L, and $\delta=0.994$ for $L=1201$~bp.

Several comments are in order. First, as expected, the asymptotic values, $L_p^\infty$ and $L_p^0$, are, at about 2~nm, the same as the ones extracted from the linear OSF and OSF--Manning fits shown in Fig.~\ref{LpOSFM}. Next, the values of the crossover ionic strength, $I_0$, is on the order of 0.1~mol/L which corresponds to a Debye screening length $\kappa_0^{-1}\simeq 1$~nm, i.e. on the order of the DNA radius, $R$. It thus suggests that the concave shape, which is more pronounced for $I\simeq I_0$ comes from non-linear Poisson-Boltzmann effects as illustrated in Section~C. Moreover, the effective power law for $I\simeq I_0$ is found by doing a logarithmic expansion of \eq{interp} around $I_0$: 
\be
\ln(L_p-L_p^\infty)\simeq \ln\left(\frac{L_p^0-L_p^\infty}2\right)-\frac{\delta}2 \ln\left(\frac{I}{I_0}\right)
\ee
which yields an exponent $-\delta/2\simeq 0.5$ which is a good approximation for $0.05<I<0.5$~mol/L (error less than 1~nm). Finally, at low ionic strength, $I\ll I_0$, \eq{interp} varies slightly and the curve looks like a linear law as a function of $I^{-1}$ with a small slope, as suggested by \eq{OSFMan}. At large Ionic strength, $I\gg I_0$, \eq{interp} yields $L_p\simeq L_p^\infty +(L_p^0- L_p^\infty) (I_0/I)^\delta$ which is equivalent to \eq{OSF} for $\delta=1$ but with a slightly larger slope of 2 to 4 nm$\cdot$L/mol.

For divalent Mg$^{2+}$ ions, the fits are also good but with different values for $\delta$ and $I_0$ as compared to the Na$^+$ only case. The parameter values $L_p^\infty$ are kept equal to the Na$^+$ case, and the values of $L_p^0$ are comparable: $L_p^0=59.2$~nm,  $I_0=0.017$~mol/L, and $\delta=0.830$ for $L=2060$~bp, and $L_p^0=51.7$~nm, $I_0=0.045$~mol/L, and $\delta=0.546$ for $L=1201$~bp. As expected the concavity is less pronounced and shifted to lower values of $I$, close to $I_0$.
The formula, \eq{interp}, can be useful for experimentalists to interpolate values of $L_p$ on the whole range of $I$.

\section{Conclusion}

\subsection{Summary}

Using the High-Throughput Tethered Particle Motion setup, we measured the impact of the ionic strength on DNA conformation for two DNA of lengths 2060~bp and 1201~bp. To this end, we investigated a large and homogeneously distributed range of ionic strengths, $I\in[0.01,3]$~mol/L, by adding salt to the buffer with monovalent Na$^+$ or divalent Mg$^{2+}$ counterions. Experimental drift and biases due to the finite exposure time of detector were were corrected. To extract properly the DNA persistence length, $L_p$, from the HT-TPM amplitude of motion, $R_{\rm exp||}$, numerical exact sampling simulations (without explicit mobile ions and solvent) were performed. Both the DNA excluded volume and the particle one's were taken into account. These simulations allowed us to obtain the experimental $L_p$ as a function of $I$ with a good accuracy of about 4\% (Fig.~\ref{Lp}). When $L_p$ is plotted as a function of $I^{-1}$, the overall trend is a monotonous increasing function with, for the Na$^+$ case, a concave shape, and, for the Mg$^{2+}$ one, an almost linear one (except at very low concentrations of Mg$^{2+}$).

Our results are compared to other results found in the literature. A quantitative comparison is difficult, since the $L_p$ values fluctuate appreciably depending on the experimental setup and the method of extraction of $L_p$. Hence, for instance at $I\simeq150$~mmol/L, $L_p$ lies between 40 and 74~nm (Table~\ref{LpPhysio}). The available experimental values can, as proposed by Savelyev \textit{et al.}~\cite{Savelyev2012}, be divided into two sets of data. The global behavior of our measured $L_p$ with $I$ is not coherent neither with the first set nor with the second set of experiments. On one hand, our $L_p$ values appear to be in agreement with the slow increase (of 10\%) of $L_p$ observed by the first group at low $I$. On the other hand, at high $I$, our $L_p$ values show a significant 25\% variation, in perfect agreement with the experiments of the second group.

Our experimental $L_p$ values follow a linear OSF law in $I^{-1}$ only for a very small range of ionic strength at high $I$ with a different prefactor than predicted by OSF in~\eq{OSF}. 

For monovalent Na$^+$ counter-ions, $L_p$ varies linearly with $I^{-1}$ at very low $I$, according to the OSF equation using the Manning counter-ion condensation theory, \eq{OSFMan}. The whole $I$ range is furthermore well fitted by \eq{LpManning}, which takes into account both the DNA internal stretching due to phosphate ions of the backbone and the counter-ion condensation around the DNA. 

For divalent Mg$^{2+}$ counter-ions, however, neither \eq{OSFMan} nor \eq{LpManning} (with the same fitting parameter as for Na$^+$) fit well the data. It suggests that these theories do not reproduce well the observed dependence with the valency $z$. Using a variational approach taking into account both non-linear Poisson-Boltzmann effects and screening by mobile ions, we propose a reasonable fit both for the Na$^+$ and the Mg$^{2+}$ cases, but only for $I<0.1$~mol/L and with two different values $L_p^0$ at vanishing ionic strength. This marked decrease of $L_p^0$ at very low $I$, when a very small amount of Mg$^{2+}$ ions is added, is semi-quantitatively explained by a theory which consider ion-ion correlations, \eq{Nguyen}, for large $z$. 

In order to both interpolate $L_p$ values in-between the ones effectively measured, and compare to future experimental data, we proposed an empirical formula which fits both the monovalent and divalent cases.

\subsection{Concluding remarks}

The large scattering of the available experimental $L_p$ values observed in Figs.~\ref{Lplit}(a) and~(b) may be due to the different experimental setups but also to the various buffers used in these experiments. Indeed, we have shown that the presence of traces of divalent ions in the buffer can decrease substantially the $L_p$ value at a given ionic strength. They are often present in buffers in order to maintain the fixed $p$H.  

An illustration of the extreme sensitivity of persistence length values to the experimental method and the model employed to extract the results was shown by Mielke \textit{et al.}~\cite{Mielke2009}. Brownian dynamics simulations on a double-stranded DNA in a bulk environment were performed at different salt concentrations. Two different strategies were employed to calculate $L_p$ from the simulation results. One used the expression  \eq{WLCcorr} of the WLC model, and the other an approximation proposed by Hagerman~\cite{Hagerman1981} of the rotational diffusion coefficients in order to directly connect to the experimental results of Ref.~\cite{Hagerman1988}. At low concentrations, depending on the method used, $L_p(I)$ splits in two distinct behaviours. Values from the rotational diffusion coefficients were  more than 30 times larger than the WLC ones and Hagerman's values. This result highlights the role of the chosen model to extract $L_p$, and the rough approximation used in earlier models to extract it from DLS measurements.

Furthermore, it has been recently shown that HT-TPM can detect the effect of the DNA sequence, in particular the presence of A-tracts, on the DNA conformation and has been interpreted as a modification of the DNA spontaneous curvature~\cite{Brunet2015}. Preliminary results also show that, for a given DNA length but two different sequences, the persistence length varies. It is thus tempting to suggest that the bare, non-electrostatic, contribution to $L_p$ can also be sequence-dependent and be another explanation for this data scattering.

Finally, many experiments study either the influence of ions with higher valency, for instance trivalent ions such as spermidine which also has a strong effect on $L_p$ at millimolar concentrations~\cite{Wang1997,Baumann1997}, and/or the role of multivalent ions on the DNA melting temperature~\cite{Owczarzy2008}. It would be interesting to pursue such a quantitative study of the DNA conformation for such trivalent ions, and thus to study the interplay between screening effects, condensation~\cite{Bloomfield1996}, and denaturation~\cite{Korolev1998}.\\

On the theoretical side, a complete theory which explains the variations of $L_p$ both as a function of $I$ and the counter-ion valency $z$ is still lacking. The approach by Manning~\cite{Manning2006} is appealing since it fits very well the experimental data for monovalent counter-ions with only one fitting parameter. However, the precise treatment of ion-ion correlations should be taken into account to extend such a theory to counter-ions with higher valencies. Note that these approaches do not consider dielectric exclusion close to a low dielectric molecule such as DNA~\cite{Chertsvy2007,Buyukdagli2010} or van der Waals interactions which are also modified when the ionic strength is varied.

In this paper we assumed that, according to the WLC model, the tangent-tangent correlation function is a simple exponential, \eq{WLCcorr}, which therefore leads to a single correlation length, $L_p$. However Barrat and Joanny show that, by taking into account the polymer bending fluctuations, the persistence length is scale-dependent~\cite{Barrat1993}. A more appropriate choice would be a double exponential where $\langle \bt(s)\cdot\bt(0)\rangle\simeq 1-s/L^\infty_p$ at small length scales and $\langle \bt(s)\cdot\bt(0)\rangle=\exp\left(-s/L^{\rm OSF}_p\right)$ at large length scales, where the crossover depends on the ionic strength. It has been shown~\cite{Manghi2004} that $L^{\rm OSF}_p$ is given by \eq{OSF} and $L_p^\infty$  is the bare persistence length. Another approach proposed that the latter is also salt-dependent~\cite{Gubarev2009}. Of course such a model is more difficult to apply to the experimental measure of $R_{\rm DNA}$ only, with a rather tricky extraction of two different correlation lengths, but it might provide a relevant framework to explain the overall observed behaviours of $L_p(I)$.

\acknowledgements
We thank Oriol Servera Sires for is valuable Bs.C. work on the fitting procedure, and Juliette Wilhem who helped for the experiments as part of her Master project. 

We acknowledge financial support from the CNRS, University of Toulouse 3 and ANR-11-NANO-010 TPM-On-a-Chip.

\end{document}